\documentclass[9pt, conference]{IEEEtran}

\IEEEoverridecommandlockouts
\usepackage{amsmath,graphicx}
\usepackage{booktabs}
\usepackage{adjustbox}
\usepackage{url}
\usepackage{amsmath,amssymb,amsfonts}
\usepackage{algorithmic}
\usepackage{graphicx}
\usepackage{textcomp}
\usepackage{xcolor}
\usepackage{url}
\usepackage{multirow,cite,bm}
\usepackage{pifont}% http://ctan.org/pkg/pifont
\newcommand{\cmark}{\ding{51}}%
\newcommand{\xmark}{\ding{55}}%
\usepackage[caption=false, position=top]{subfig}

\def\BibTeX{{\rm B\kern-.05em{\sc i\kern-.025em b}\kern-.08em
    T\kern-.1667em\lower.7ex\hbox{E}\kern-.125emX}}
\begin{document}

\title{Enhancing Listened Speech Decoding from EEG via \\ Parallel Phoneme Sequence Prediction
\thanks{This project was partially supported by a USC Annenberg Graduate Fellowship and also by the Defense Advanced Research Projects Agency (DARPA) under cooperative agreement No. N660012324006. The content of the information does not necessarily reflect the position or the policy of the Government, and no official endorsement should be inferred.}
}

\author{Jihwan Lee, Tiantian Feng, Aditya Kommineni, Sudarsana Reddy Kadiri, Shrikanth Narayanan \\ \\
\IEEEauthorblockA{Signal Analysis and Interpretation Laboratory, University of Southern California, Los Angeles, USA}}
% jihwan@usc.edu}

\maketitle

\begin{abstract}
Brain-computer interfaces (BCI) offer numerous human-centered application possibilities, particularly affecting people with neurological disorders. % to those with communication disorders.
Text or speech decoding from brain activities is a relevant domain that could augment the quality of life for people with impaired speech perception.
We propose a novel approach to enhance listened speech decoding from electroencephalography (EEG) signals by utilizing an auxiliary phoneme predictor that simultaneously decodes textual phoneme sequences.
The proposed model architecture consists of three main parts: EEG module, speech module, and phoneme predictor.
The EEG module learns to properly represent EEG signals into EEG embeddings.
The speech module generates speech waveforms from the EEG embeddings.
The phoneme predictor outputs the decoded phoneme sequences in text modality.
Our proposed approach allows users to obtain decoded listened speech from EEG signals in both modalities (speech waveforms and textual phoneme sequences) simultaneously, eliminating the need for a concatenated sequential pipeline for each modality. 
The proposed approach also outperforms previous methods in both modalities.
The source code and speech samples are publicly available\footnote{\url{https://github.com/lee-jhwn/icassp25-fesde-phoneme}}.

\end{abstract}


\section{Introduction}
\label{sec:intro}

Brain-computer interfaces (BCI) offer a promising avenue for a wide range of applications, particularly in assisting individuals with neurological and communication disorders.
The ability to decode perceived, imagined, or produced speech, as well as text, from brain signals is fundamental to the development and effectiveness of these technologies.
The successful integration of speech and text decoding into BCI leads to significant advances in neurorehabilitation, offering alternative communication means \cite{willett2023high, metzger2023high, puffay2023relating}.

% Several 
% Two types of signal - invasive and non-invasive
% There are mainly two ways of measuring brain signals: invasive and non-invasive. The invasive measurements, such as ECoG, promise more clear recordings of the signals, however,  measurement methods are Commonly used recording types of brain signals are ECoG and EEG

% Non-invasive measurements, such as EEG, are less costly and more economical, compared to invasive measurements such as ECoG, as they do not require any surgical step.
% However, modeling the recorded signals is quite challenging due to its low signal-to-noise ratio, as the brain signals are recorded outside the scalp, rather than inside.

% Aditya to fill up EEG SSL paragraph 
\par 
Recent advances in self-supervised learning 
% has showed remarkable performances in the domains of speech, audio and vision. These methods 
show promise for processing biosignals such as EEG that tend to be inherently noisy. This has led to development of pre-trained models to learn broadly meaningful representations that can be generalized to various downstream tasks. %enabling generalizability to downstream tasks. 
Recent studies \cite{kommineni2024knowledgeguided, kostas2021bendr, chien2022maeeg} have explored encoder-decoder architectures showing the ability to capture representations of the EEG signal that can generalize across various domains.such as Motor-Imagery, Sleep and other Event Related Tasks.

Several methods have been proposed to decode text from EEG signals in the context of human speech and language processing. These include methods that 
%They propose to 
utilize a pre-trained language model and the transformer architecture \cite{wang2022open, eeg2text}, or the quantized variational encoder \cite{duan2024dewave}.
Liu et al. \cite{eeg2text} propose to decode textual information in the EEG signals by adopting the transfer learning regime on the transformer architecture and using a pre-trained language model.

Various approaches have been proposed to decode acoustic speech information from EEG signals. They involve using convolutional neural networks (CNN) \cite{vlaai, xu2024convconcatnet, bras24_is}, pre-trained speech models \cite{meta-paper}, or the WaveNet \cite{vandenoord16wavenet} architecture \cite{fangyuan24}. 
Recently, a framework for direct reconstruction of listened speech waveforms has been proposed as described in \cite{fesde}, where no intermediate acoustic feature step is required.

There are instances where the need arises to decode speech and text from brain signals simultaneously.
For example, in the case of real-time BCI communication scenarios involving hearing or speech disabilities, simultaneous decoding in both modalities can be beneficial. 
However, most existing approaches are limited to only a single modality at a time.
Consequently, to achieve decoding in both modalities, the conventional methods require implementing a concatenated, sequential pipeline, such as an additional speech synthesizer or recognizer.
% Hence, a concatenated, sequential pipeline followed by an additional speech synthesizer or speech recognizer is required to acquire decoded speech in both modalities.
% and the sequential decoding of the other followed by, such as an additional speech synthesizer followed by decoded text, or a speech recognizer followed by decoded speech.

We propose a novel framework that can decode listened speech in both modalities (speech waveform and textual phoneme sequences) simultaneously by incorporating an auxiliary phoneme predictor.
The proposed framework consists of three main parts: EEG module, speech module, and phoneme predictor.
The EEG module focuses on learning EEG embedding representation from the EEG signals.
The speech module aims to generate listened speech waveform from the EEG embeddings.
The phoneme predictor decodes the phoneme sequences from the EEG embeddings.
The proposed framework not only allows parallel decoding of EEG signals into both %of the modalities of 
speech waveforms and phoneme sequences, but also outperforms the previous methods in listened speech decoding. 
Table~\ref{tab:compare-model} shows the comparison of the proposed model with existing models in decoding speech waveforms and phoneme sequences.

\begin{table}[t]
    % \vspace{0.2 cm}
% \tiny
  \caption{Comparison of proposed model with existing models in decoding speech waveform and phoneme sequences.}
    \vspace{-0.1cm}
  \label{tab:compare-model}
  \centering
  \begin{tabular}{lcc}
    \toprule
    \multicolumn{1}{l}{\textbf{Model}} & \multicolumn{2}{c}{\textbf{Can it decode ... ?}} \\
     & \multicolumn{1}{c}{Speech Waveform} & \multicolumn{1}{c}{Phoneme Sequence}\\
     % \multicolumn{2}{c|}{}&unseen audio&unseen subject&unseen both&unseen audio&unseen subject&unseen both\\
    \toprule
\textsc{FESDE \cite{fesde}} &\cmark & \xmark \\
 \textsc{EEG2Text \cite{eeg2text}}&  \xmark & \cmark \\
 \midrule
  \textsc{Ours}&  \cmark & \cmark \\

    \bottomrule
  \end{tabular}
      \vspace{-4mm}
\end{table}

% The proposed framework utilizes an auxiliary phoneme predictor 

Our main contributions are as follows:
\begin{itemize}
\item We introduce the novel framework that enables parallel decoding of listened speech waveforms and phoneme sequences directly from EEG signals. To the best of our knowledge, this is the first approach to achieve parallel decoding in this setting.

\item We demonstrate that the incorporation of an auxiliary phoneme predictor enhances the performance of listened speech decoding from EEG signals, outperforming previous approaches.

\item We provide a phoneme-level analysis of the model's ability to decode phoneme sequences and speech waveforms.
\end{itemize}

\section{Proposed Method}

\subsection{Model Architecture}
Our proposed framework consists of three major parts: the EEG module, the speech module, and the phoneme predictor. The EEG module learns the latent representation embeddings from EEG signals. The EEG embeddings are fed into the phoneme predictor and the speech module in parallel to output the predicted phonemes and decoded speech waveforms, respectively. % corresponding to given EEG signals, respectively.
Note that our approach enables parallel decoding of both speech waveforms and phoneme sequences, rather than requiring a sequentially concatenated pipeline for each modality.
For detailed implementation, refer to the source code provided.

\begin{figure}[t]
  \centering
  \includegraphics[width=\linewidth,trim={1.1cm 0.1cm 0.4cm 0},clip]{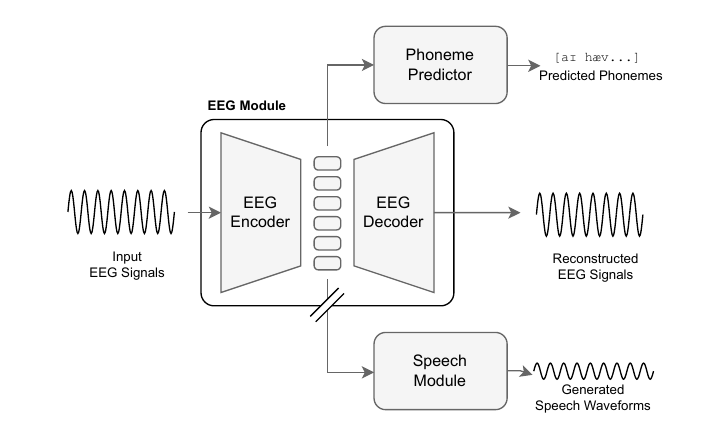}
  \caption{The overall architecture of the proposed framework. The EEG module learns the EEG embeddings, which are then fed in parallel to both the phoneme predictor and the speech module to decode phoneme sequences and speech waveforms simultaneously.}
  \label{fig:overall-archi}
\end{figure}

\subsubsection{EEG and Speech Module}
We adopt the EEG and speech module from \cite{fesde}. The EEG module consists of an EEG encoder and an EEG decoder, where it learns to represent input EEG signals into EEG embeddings.
The EEG encoder comprises of convolution blocks followed by a Structured State Space Sequence (S4) \cite{s4_paper} block, as in \cite{fesde, kommineni2024knowledgeguided}.
The EEG decoder is composed of 1D transpose convolution blocks.
The EEG encoder takes preprocessed EEG signals as input and ouputs EEG embeddings, whereas the EEG decoder outputs reconstructed EEG signals from the EEG embeddings.

The speech module learns to generate the speech waveform from EEG embeddings, similar to \cite{fesde, vits}.
It consists of a speech encoder, a speech decoder, and a connector.
As in \cite{fesde,vits}, the speech encoder comprises of non-causal WaveNet \cite{vandenoord16wavenet} residual blocks followed by a projection layer, and the speech decoder is HiFi-GAN V1 \cite{hifigan}.
The speech encoder takes linear spectrograms of speech as input and outputs speech embeddings. The speech decoder then learns to generate speech waveforms from the speech embedding.
The EEG embeddings are converted to speech embeddings through the connector, consisting of a transformer encoder \cite{transformer} followed by a projection layer and the normalizing flow network, as in \cite{fesde, vits}.
The normalizing flow network is composed of affine coupling layers \cite{dinh2017density}, which act as an invertible function between EEG and speech embeddings, aligning these two distributions.
A gradient stop is applied between the EEG module and the speech module, hence the training of the EEG module is not influenced by the speech module.

\subsubsection{Phoneme Predictor}
The phoneme predictor consists of conformer blocks \cite{conformer} followed by a phoneme decoder, which is widely recognized in the automatic speech recognition (ASR) domain.
The phoneme decoder is a single layer of LSTM with an attention mechanism.
The phoneme predictor takes the EEG embeddings as input and outputs the phoneme sequences.
This process allows the phoneme information to be more precisely captured in EEG embeddings.
% to better capture the phoneme information.

\subsection{Training Objectives}
We use the connectionist temporal classification (CTC) loss \cite{ctc-loss} for the phoneme predictor, as defined in Eq.~(\ref{equation:ctc}).
The CTC loss is widely employed in the ASR domain due to its ability to effectively manage many-to-one sequence mapping.
\begin{align}
L_{\text{CTC}}(z, \hat{z}) = - log\sum_{\pi \in B^{-1}(z)}P(\pi | \hat{z})
  \label{equation:ctc}
\end{align}
where $z$ and $\hat{z}$ are the target and predicted phoneme sequences, respectively, and $B^{-1}$ indicates the set of possible sequences that can collapse into $z$.

For the EEG module, we use the reconstruction loss for EEG signals, identical to \cite{fesde, kommineni2024knowledgeguided}, as defined in Eq.~(\ref{equation:loss-cossim}):
\begin{align}
L_{\text{EEG}}(x, \hat{x}) = 1 - \frac{1}{N_{\text{ch}}}\sum_{i=1}^{N_{\text{ch}}}\frac{x_i^T \cdot \hat{x_i}}{\Vert x_i \Vert \Vert \hat{x_i}\Vert}
  \label{equation:loss-cossim}
\end{align}
where $N_{\text{ch}}$ is the number of EEG channels ($N_{\text{ch}}=128$), and $x_i$ and $\hat{x_i}$ are $i\textsuperscript{th}$ channel of the input and the reconstructed EEG signals, respectively.
The total training objective of the phoneme predictor and the EEG module is as in Eq.~(\ref{equation:loss-pheeg}):
\begin{align}
L=L_{\text{EEG}} + \alpha L_{\text{ctc}}
\label{equation:loss-pheeg}
\end{align}
We heuristically chose $\alpha=0.3$.

To train the speech module, we adopt the training objectives from \cite{fesde,vits}, which consist of the mel-spectrogram reconstruction loss, the KL divergence loss between the latent representations of EEG and speech, and the GAN loss for the speech waveform generation. 
A separate optimizer was used to train the speech module, as a stop gradient is applied between the EEG module and the speech module.

\begin{table*}[htb!]
\footnotesize
  \caption{Speech decodability evaluation: MCD (dB) and Mel-Corr (\%) for each configuration, with 95\% confidence intervals. Lower MCD and higher Mel-Corr values indicate better performance. For convenience, the Mel-Corr values are scaled by a factor of 100.}
  
  \label{tab:mcd-corr}
  \vspace{-0.1cm}
  \centering
  \begin{tabular}{cc|ccc|ccc}
    \toprule
    \multicolumn{2}{c|}{\multirow{2}{*}{\textbf{Model}}} & \multicolumn{3}{c|}{\textbf{MCD (dB) $\downarrow$}} & \multicolumn{3}{c}{\textbf{Mel-Corr (\%) $\uparrow$}}\\
     \multicolumn{2}{c|}{}&unseen speech&unseen subject&unseen both&unseen speech&unseen subject&unseen both\\
    \toprule
    \multicolumn{2}{c|}{\textsc{FESDE \cite{fesde}}} & $11.84 \pm 0.12$ & $11.76 \pm 0.11$ & $11.65 \pm 0.27$ & $13.97 \pm 0.70$ & $13.65 \pm 0.67$ & $13.05 \pm 1.99$ \\
    \midrule
    \multirow{3}{*}{\textsc{Ours}} & \textsc{CB-0} & $\bm{10.18 \pm 0.11}$ & $\bm{9.58 \pm 0.16}$ & $\bm{10.29 \pm 0.35}$ & $\bm{27.10 \pm 1.05}$ & $\bm{32.03 \pm 1.48}$ & $\bm{28.34 \pm 3.17}$ \\
     & \textsc{CB-1} & $10.43 \pm 0.11$ & $10.05 \pm 0.12$ & $\bm{10.30 \pm 0.32}$ & $\bm{26.37 \pm 0.94}$ & $28.46 \pm 1.06$ & $25.99 \pm 2.66$ \\
  & \textsc{CB-2} & $11.21 \pm 0.12$ & $11.11 \pm 0.12$ & $11.28 \pm 0.35$ & $22.33 \pm 0.78$ & $21.97 \pm 0.82$ & $22.20 \pm 2.47$ \\

    \bottomrule
  \end{tabular}
  
\end{table*}
\begin{table*}[ht]
  \caption{Phoneme sequence decodability evaluation: Top-k accuracy (\%) of subsequent phoneme prediction given previous phonemes, with 95\% confidence intervals.}
    \vspace{-0.1cm}

\begin{adjustbox}{width=1\textwidth}
% \tiny
  \label{tab:topk}
  % \hspace(\pagewidth)
  \centering
  \begin{tabular}{cc|ccc|ccc|ccc}
    \toprule
    \multicolumn{2}{c|}{\multirow{2}{*}{\textbf{Model}}} & \multicolumn{3}{c|}{\textbf{Top-1 Accuracy (\%) $\uparrow$}} & \multicolumn{3}{c|}{\textbf{Top-3 Accuracy (\%) $\uparrow$}} & \multicolumn{3}{c}{\textbf{Top-5 Accuracy (\%) $\uparrow$}}\\
     \multicolumn{2}{c|}{}&unseen speech&unseen subject&unseen both&unseen speech&unseen subject&unseen both&unseen speech&unseen subject&unseen both\\
    \toprule
    \multirow{2}{*}{\textsc{EEG2Text \cite{eeg2text}}} & \textsc{sc} & $43.12 \pm 0.80$ & $50.17 \pm 0.75$ & $43.08 \pm 2.36$ & $58.95 \pm 0.81$ & $69.58 \pm 0.65$ & $58.73 \pm 2.40$ & $66.65 \pm 0.84$ & $79.05 \pm 0.59$ & $66.38 \pm 2.55$ \\
    & \textsc{pt} & $42.96 \pm 0.80$ & $53.38 \pm 0.74$ & $42.67 \pm 2.36$ & $61.02 \pm 0.78$ & $71.51 \pm 0.65$ & $60.84 \pm 2.39$ & $69.72 \pm 0.83$& $80.84 \pm 0.57$& $69.70 \pm 2.52$\\
    \midrule
    \multirow{3}{*}{\textsc{Ours}} & \textsc{CB-0} & $33.17 \pm 0.98$ & $\bm{59.04 \pm 1.05}$ & $33.65 \pm 2.91$ & $55.12 \pm 1.16$ & $\bm{83.10 \pm 0.82}$ & $55.45 \pm 3.47$ & $65.81 \pm 1.17$ & $\bm{91.17 \pm 0.60}$ & $66.01 \pm 3.56$ \\
     & \textsc{CB-1} & $41.56 \pm 1.23$ & $51.07 \pm 1.19$ & $41.02 \pm 3.82$ & $58.00 \pm 1.25$ & $68.16 \pm 1.08$ & $58.14 \pm 3.76$ & $65.47 \pm 1.20$ & $75.80 \pm 0.92$ & $65.92 \pm 3.51$ \\
  & \textsc{CB-2} & $\bm{49.45 \pm 0.89}$ & $\bm{59.73 \pm 0.64}$ & $\bm{49.72 \pm 2.59}$ & $\bm{65.81 \pm 0.84}$ & $78.50 \pm 0.58$ & $\bm{65.80 \pm 2.49}$ & $\bm{73.72 \pm 0.82}$ & $85.81 \pm 0.51$& $\bm{73.69 \pm 2.40}$ \\

    \bottomrule
  \end{tabular}
\end{adjustbox}
\end{table*}

\section{Experiments}
\subsection{Dataset}
Adopting the experimental setup of \cite{fesde}, we conducted our experiments on the same N400 dataset \cite{n400}.
The 128-channel EEG signals were collected from 24 subjects at a sampling rate of 512 Hz. Each subject listened to 440 synthesized, gender-neutral English speech utterances.
We chose two subjects and 40 sentences as held-out test sets, yielding three different test sets.
% : \texttt{unseen speech}, \texttt{unseen subject}, and \texttt{unseen both}.
In the \texttt{unseen speech} or \texttt{unseen subject} test set, only the speech sentences or the subjects are held-out, respectively. In the \texttt{unseen both} test set, both speech sentences and subjects are held-out.
% There are three test sets, unseen speech, unseen

We employ the same EEG pre-processing pipeline as described in \cite{fesde}, which includes powerline noise removal using a notch filter at 60 Hz, preservation of EEG spectral information through a bandpass filter (0.5–50 Hz), eye blink artifact removal via independent component analysis (ICA), and downsampling the EEG data to 256 Hz. %: powerline removal (notch filter at 60 Hz), EEG spectral information preservation (bandpass filter with low of 0.5 Hz and high 50 Hz), eye blink removal (using independent component analysis), and downsampling to 256 Hz.
% Following the audio configuration setup in \cite{vits, fesde},
The speech samples were downsampled to 22,050 Hz, as described in \cite{vits, fesde}.
%The following short-time Fourier transformation configuration was utilized for spectrograms: a window and FFT size of $1024$ with a hop size of $256$ and $80$ mel-bands. 
The mel-spectrograms were generated using a short-time Fourier transform (STFT) with a window and FFT size of 1024, a hop size of 256, and 80 mel-frequency bands.
We use \texttt{espeak G2P}\footnote{\url{https://github.com/espeak-ng/espeak-ng}} to convert the raw text (graphemes) to phonemes.

\subsection{Experimental Setup}

We compare three different configurations of our proposed framework. In \textsc{CB-1} and \textsc{CB-2}, the phoneme predictor utilizes one and two conformer blocks, respectively. In contrast, \textsc{CB-0} does not use any conformer blocks; instead, it employs a single LSTM layer with an attention mechanism.

We compare our proposed model against the following baselines: \textsc{FESDE} \cite{fesde} for speech decoding, and \textsc{EEG2Text} \cite{eeg2text} for phoneme sequence decoding.
To better align \textsc{EEG2Text} with our task and ensure a fair comparison, we made a few modifications. The pre-trained language model was removed, as it had been originally designed to predict text tokens rather than phoneme sequences. Instead, we replaced it with a single LSTM layer combined with an attention mechanism, identical to the phoneme decoder used in our proposed model. 
We compared two configurations of \textsc{EEG2Text} with our approach: \textsc{EEG2Text-sc}, which was trained from scratch without any pre-trained initial weights, and \textsc{EEG2Text-pt}, which was first pre-trained using EEG input alone, followed by transfer learning for the phoneme prediction task. In \textsc{EEG2Text-pt}, the EEG reconstruction loss, as defined in Eq.~(\ref{equation:loss-cossim}), was applied during the pre-training phase. % rather than the mask prediction loss. in order to minimize the number of different factors for more simple comparison.

All models were trained using AdamW optimizer \cite{loshchilov2018decoupled} with a learning rate of $0.0002$. Due to the stop gradient between the EEG module and the speech module, a separate optimizer was used for the speech module. All of the training configurations of the proposed methods and \textsc{FESDE} \cite{fesde} were trained for 100k iterations. The \textsc{EEG2Text-sc} configuration was trained for 70k iterations. For \textsc{EEG2Text-pt}, pre-training was conducted for 150k iterations using the EEG data only, followed by 70k iterations for the phoneme sequence prediction task. 
%The \textsc{EEG2Text-pt} configuration was pre-trained for 150k iterations using only the EEG data, then 70k iterations for the phoneme prediction task. 
One Nvidia A40 GPU was utilized for each training configuration.

\begin{figure*}[!ht]
  \centering
  \subfloat[MCD (dB) of each consonant (bluish) and vowel (reddish) group.]{%
  \includegraphics[clip, width=0.98\textwidth,height=4.2cm]
  {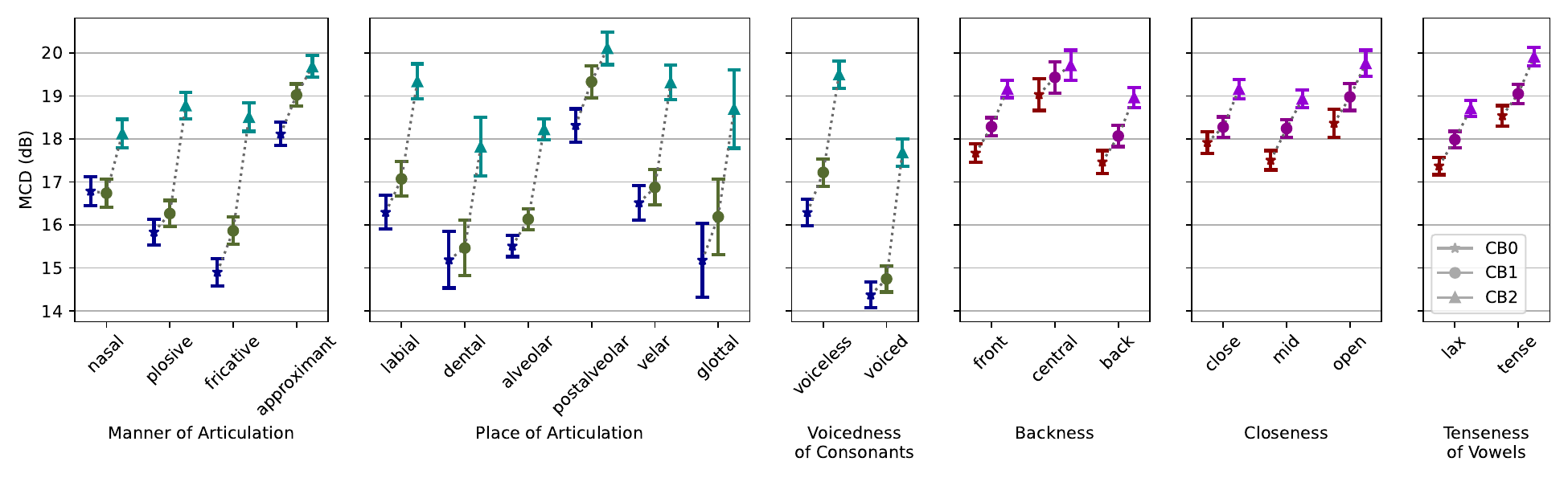}%
  }
  \vspace{-3mm}

  \subfloat[Mel-Corr (\%) of each consonant (bluish) and vowel (reddish) group.]{%
  \includegraphics[clip, width=0.98\textwidth,height=4.2cm]
  {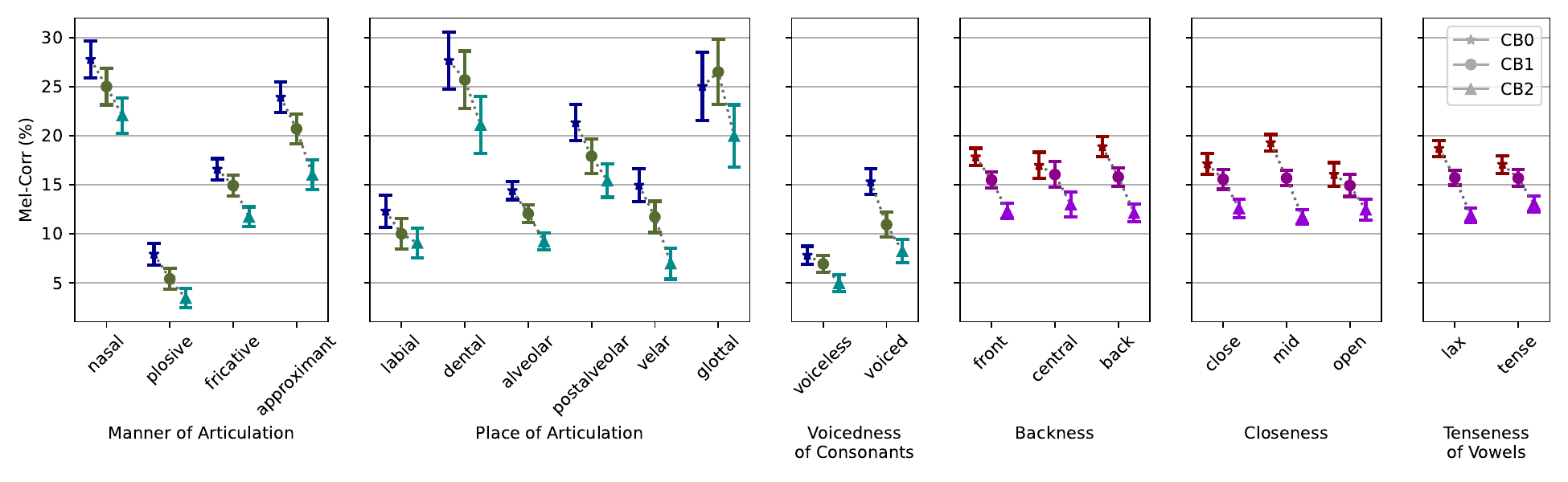}%
  }

  \vspace{-3mm}

  \subfloat[Top 3 Accuracy (\%) of each consonant (blue) and vowel (red) group.]{%
  \includegraphics[clip, width=0.98\textwidth,height=4.2cm]
  {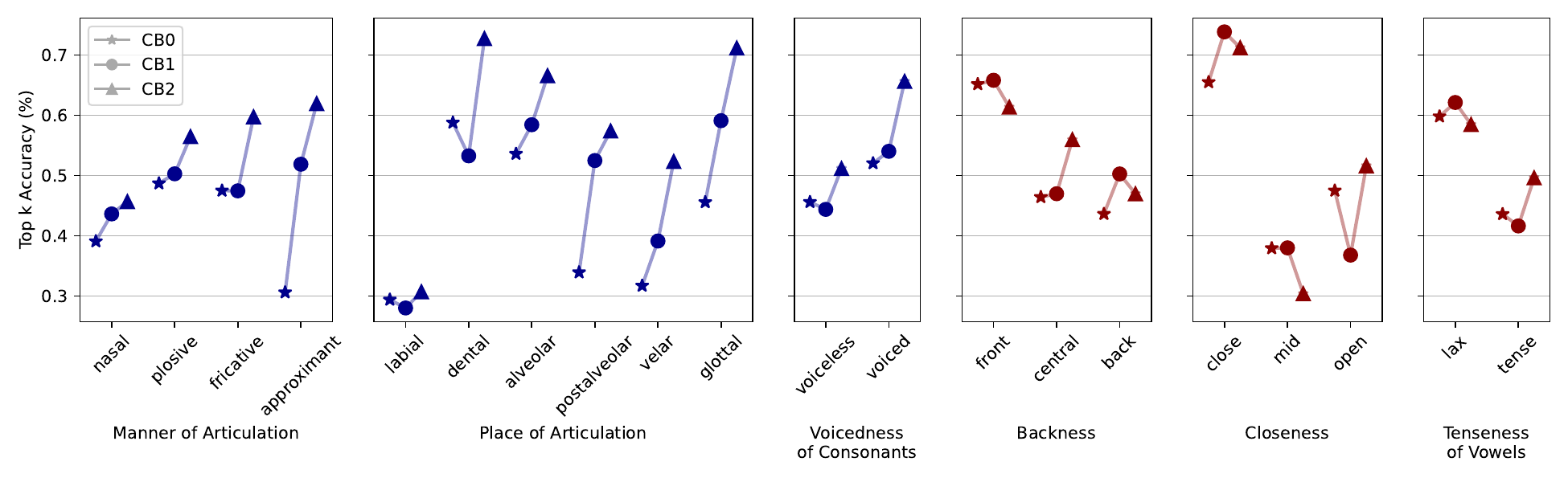}%
  \label{fig:phonemes-top3}
  }

  \vspace{-1mm}

\caption{(a) MCD (dB), (b) Mel-Corr (\%), and (c) Top-3 accuracy (\%) of each phoneme group with respect to the number of the conformer blocks. Better performance is indicated by the lower MCD, and higher Mel-Corr and top-3 accuracy. The consonants (bluish colors) are grouped by their manner, place, and voicedness of articulation. The vowels (reddish colors) are grouped by their tongue positions and tenseness.}
  \label{fig:phonemes}
\end{figure*}

\section{Results and Discussion}
\subsection{Evaluation of Speech Decoding}
We evaluate our results on the two modalities of model outputs: audio speech and phoneme sequences. 
For evaluating speech, we adopt the evaluation measures used in \cite{fesde}: mel-cepstral distortion (MCD) \cite{mcd} and the Pearson correlation between two mel-spectrograms (Mel-Corr). MCD quantifies the Euclidean distance between two mel-cepstral coefficients (MCC):
\begin{align}
\vspace{-0.5cm}
\text{MCD} = \frac{10\sqrt{2}}{\ln10}\sqrt{\sum_{i=1}^{N_\text{MCC}}(\text{MCC}_i-\widehat{\text{MCC}_i})^2}
  \label{equation:mcd}
  \vspace{-0.5cm}
\end{align}
Lower MCD values and higher Mel-Corr values indicate higher quality in the decoded speech waveforms. For simplicity, Mel-Corr values are reported after being multiplied by 100.
As shown in Table~\ref{tab:mcd-corr}, all of our training configurations outperform the baseline \cite{fesde} in both MCD and Mel-Corr metrics.
Our proposed approach achieves noticeable improvement, particularly in Mel-Corr, where the metrics are nearly doubled.
The best performance is achieved when a single LSTM layer is used as the phoneme predictor.
This may be attributed to the model's overemphasis on the phoneme sequence prediction task due to the increased complexity of the phoneme predictor architecture, however, further investigation is required to provide a more definitive explanation.
% THIS COULD BE DUE TO....

\subsection{Evaluation of Phoneme Sequence Decoding}
To assess phoneme sequence decoding, we compute the top-k accuracy of subsequent phoneme prediction given the EEG signals and previous phoneme inputs.
We find the top-k accuracy to be more reliable than the raw phoneme error rates (PER) because, in our experimental setup, there is no external language model to correct errors. Consequently, initial errors can lead to significant error accumulation.
As shown in Table~\ref{tab:topk}, our proposed approach outperforms both training configurations of the baseline \cite{eeg2text}, when two conformer blocks are used in the phoneme predictor.
The high accuracy in the \texttt{unseen subject} test set may be due to overfitting, as the target phoneme sequences are already seen during training in this test set. 

\subsection{Effects of Conformer}

We observe a trade-off in performances across modalities as the number of conformer blocks in the phoneme predictor increases.
Specifically, speech decodability tends to decrease as more conformer blocks are used, whereas phoneme sequence decodability shows the opposite trend. Despite this, speech decoding consistently outperforms the baseline \cite{fesde}, even in the most challenging test conditions.
We further analyze the characteristics of the phonemes to understand where this performance trade-off happens in the next Section~\ref{sec:phoneme-analysis}.

\subsection{Analysis on Phoneme Groups}
\label{sec:phoneme-analysis}
Herein, we aim to identify specific phoneme groups that perform poorly/well on each modality (1) in general; and (2) with respect to the number of the conformer blocks.
Figure~\ref{fig:phonemes} depicts the decodability of each phoneme group in relation to the number of conformer blocks. We used the same analysis pipeline as in \cite{fesde}, where the Montreal Forced Aligner \cite{mfa} was employed for phoneme segmentation. The intervals of the ground-truth phonemes are assumed to be identical to those of the reconstructed ones.
Consonants are grouped by manner, place, and voicedness of articulation, while vowels are grouped by tongue position and tenseness.
The \texttt{unseen subject} test set is excluded from the top-3 accuracy analysis in Figure~\ref{fig:phonemes-top3}, as the phoneme sequences are already seen during training and it shows a different tendency compared to the other test sets.

In general, nasal, dental, and voiced consonants are more easily decoded in the speech modality. For phoneme sequences, nasal, labial, and voiceless consonants are more challenging to decode. Front, close, and lax vowels also tend to be more easily decoded in the phoneme sequence modality.

With respect to the number of the conformer blocks, the performance trade-off between the modalities is more pronounced for consonants than for vowels.
In the case of speech modality, the performance drops more significantly for plosive, fricative, dental, velar, and voiced consonants, as the number of conformer blocks increases. Conversely, in terms of phoneme sequence decodability, the top-3 accuracy increases on most of the consonant groups except for the labial consonants.

% \section{Ablation Study}
% \subsection{EEG Pre-training}
% Contrary to \cite{fesde,eeg2text}, we observe performance drop in our proposed framework, when the EEG module is initialized with pre-trained parameter weights.

\section{Conclusion and Future Work}
We propose a framework to decode listened speech from EEG signals utilizing an auxiliary phoneme predictor, enhancing its performance in both speech and textual phoneme sequence decoding.
The proposed framework enables parallel decoding of both speech and phoneme sequences, eliminating the need for a concatenated, sequential pipeline for each modality decoder.
In the future, we plan to expand our work to speech production tasks, such as decoding phonated, attempted, or imagined speech from neural signals, as our current work is limited to just the speech perception task. Additionally, we plan to further improve the performance by leveraging pre-trained models for each modality.

% \clearpage
\bibliographystyle{IEEEtran}
\bibliography{refs}

\end{document}